# OFDM Transmission Performance Evaluation in V2X Communication

Aymen SASSI, Faiza CHARFI, Lotfi KAMOUN,
Yassin ELHILLALI, Atika RIVENQ

*The Vehicle to Vehicle and Vehicle to Infrastructure V2X communication systems are one of the main topics in research domain. Its performance evaluation is an important step before their on board's integration into vehicles and its probable real deployment. Being the first of a series of three publications, this paper studies the physical layer PHY of the upcoming vehicular communication standard IEEE 802.11p. An IEEE.802.11p PHY model, with much associated phenomena, is implemented in Vehicle to Vehicle V2V and Vehicle to Roadside unit V2I, situations through different scenarios. The series of simulation results carried out, perform data exchange between high-speed vehicles over different channels models and different transmitted packet size. We underline several propagation channel and other important parameters which affect both the physical layer network performance and the quality of transmission QoT. The Bit Error Rate BER versus Signal to Noise Ratio SNR of all coding rates is used to evaluate the performance of the communication.*

## I. Introduction

A huge quantitative as well as qualitative transport development has been achieved during the last decade. Vehicles are becoming increasingly intelligent. They are able to detect the probable hazardous obstacles around. Also, a new need for V2X sharing information is emerging. The V2X communication system is an important integral part of the Intelligent Transport Systems ITS future architecture. The associated safety applications such as driving assistance, traffic efficiency enhancement applications and commercial applications are a matter of huge topic. The IEEE community is working on a freshly standardized technology devoted to ITS communication, namely IEEE 802.11p.

The IEEE 802.11p, also known as Wireless Access in the Vehicular Environment WAVE, is an expansion to the IEEE 802.11 protocol that adds wireless communication systems in a mobile environment. It uses the mechanism initially provided by IEEE 802.11 to operate in the Dedicated Short Range Communication DSRC. This standard provides both the physical layer PHY specifications for the DSRC and the Medium Access Control MAC of vehicular communication.

In this paper, we perform the V2X PHY layer model using grouping models over different propagation environments. Several simulations are carried out using the PHY layer proposal for 802.11p to analyze different Orthogonal Frequency Division Multiplexing OFDM performance according to the nodes mobility, transmission channel type and different packet size.

The structure of this paper is as follows: section 2 deals with the state of art for the IEEE PHY layer, section 3 gives a brief overview of the 802.11p specifications, section 4 presents the transmitting process system architecture, the General Analysis and simulation results are shown in Section 5 and in Section 6, conclusions are drawn and future works introduced.

## II. State of art:

The study of the V2X communication system performance is an important preliminary step before its integration into vehicles. Numerous studies are actively being conducted for this purpose. Different scenarios in various modulation types have been simulated in V2V transmission in [1], [2]. An overview on DSRC technology and 802.11p system is given in [3] and [4]. In [5] the author studied the influence of the propagation channel on V2X physical layer performance. He gave a DSRC technology description of V2V and V2I. In [6] Vegni and Little described a hybrid vehicular communication protocol and the mechanism by which a message can be propagated under this technique. Resta, Santi and Simon studied in [7] the probability that a vehicle could correctly receive a message within a fixed time interval. In [8] the authors presented a framework to evaluate the PHY layer of 802.11p system. They tried to details the design and the implementation of an FPGA- based real-time vehicular channel emulator. Bernado, Czink, Zemen and Belanovie in [9] used a vehicular non-stationary channel model to implement the IEEE 802.11p PHY layer. In [10] the authors developed a system to use antenna steering to optimize the radio performance when connecting to a roadside Access Point AP. They demonstrated that significant gains can be achieved in the throughput by improving connectivity duration and SNR.

Matolak and Sen, provided in [11] a channel modeling results based upon measurements of V2V mobile channel, taken in the 5 GHz frequency band.

In this work, we try to build a variety of simulations to apprehend the behavior of this standard and to evaluate the transmission performance of V2X communication according to the variance of speed nodes, channel type and transmitted packet size. The following sections will introduce the different specifications of 802.11p and subsequently the simulation performed.

### III. 802.11p PHY specifications:

This section is an overview of the 802.11p standard. Indeed, it shows the architecture of the Physical Layer Management of this standard. It also presents the WAVE frame format and its different specifications.

1. Physical Layer Management :

The physical layer represents an interface between the MAC layer and the support that allows sending and receiving frames [13]. The physical layer of the IEEE 802.11p is similar to that of IEEE 802.11a. It is composed of two sub layers Fig.1. The first one is the PLCP sub-layer (Physical Layer Convergence Protocol) which is responsible for communicating with the MAC layer. It is also a convergence process that transforms the PDU (Packet Data Unit) arriving from the MAC layer to form an OFDM frame. The second is the PMD sub-layer (Physical Medium Access) which is the interface to the physical transmission medium (radio channel). Its role is to manage the encoding of data and perform the modulation

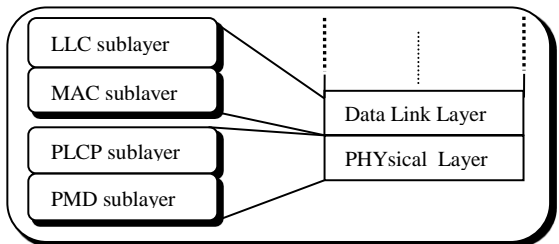

Fig. 2: 802.11p PHY Layer Management.

2. PHY Specification of the 802.11p

The IEEE 802.11p, also known as WAVE, is an expansion to the IEEE 802.11 protocol that adds wireless communication systems in a mobile environment [11]. It uses the mechanism initially provided by IEEE 802.11 to operate in the DSRC which is a communication technology based on IEEE 802.11a to work in the 5.9 GHz band (U.S.) or 5.8 GHz band (Japan, Europe). It offers data-exchange between vehicles V2V and between vehicles and roadside infrastructure V2I within a range of length reaching over 1000 m with a transmission rate of 3Mbps to 27Mbps and a transmission speed more than 260 km/h. This type of wireless technology operates on 9-channels. Each of these 9 channels is allocated a frequency band Fig. 2. Channel CH178-5.890 GHz is a control channel. Its role is to establish communications and broadcast transmissions. CH172-5.860 GHz and CH184-5.920 GHz are both safety-dedicated channels. The first one provides a critical security while the second plays a protective role against congestion on service channels. The six other service channels are allocated for bidirectional association between different types of units such as the On Board Units OBU and the Road Side Units RSU.

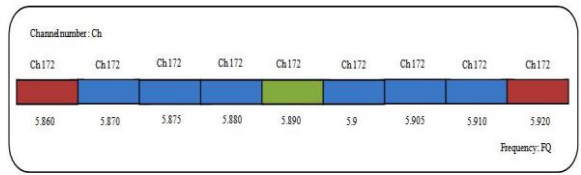

Fig. 1 : 802.11p Functional Channels.

The 802.11p channel bandwidth is halved, to adapt the standard to vehicular communication, resulting in a 10 MHz bandwidth instead of 20 MHz for 802.11a. The carrier spacing is reduced by half (0.15625 MHz) compared to 802.11a (0.3125 MHz) while the symbol length for 802.11p is twice that of 802.11a (4.0μs) Table.1 [12], [13]. Consequently, the transmission rate is reduced by half. Various modulation schemes (BPSK, QPSK, 16 QAM and 64 QAM) are used for efficient packet transmission. The pilot signals are inserted into the subcarriers of -21, -7, 7 and 21 which are then processed by an Inverse Discrete Fourier Transform (IDFT) modulation and transmitted after a Cyclic Prefix (CP) is added. In the receiver unit, after timing synchronization, the CP is removed before the signal passes through discrete Fourier transform (DFT) demodulation.

| Parameter | 802.11p | 802.11a |
|---|---|---|
| $N_{SD}$: Nber of sub-carrier | 48 | 48 |
| $N_{SP}$: Nber of pilot sub-carrier | 4 | 4 |
| $N_{ST}$: Nber of sub-carrier Total | 52 (NSD + NSP) | 52 (NSD + NSP) |
| $N_{STS}$: Nber of sub-carrier STS | 12 | 12 |
| $T_{STS}$: STS duration | 1.6 µs ($T_{FFT}/4$) | 0.8 µs ($T_{FFT}/4$) |
| $T_{Short}$: Short training sequence duration | 16 µs (10 × $T_{STS}$) | 8 µs (10 × $T_{STS}$) |
| $N_{LTS}$: Nber of sub-carrier LTS | 53 | 53 |
| $T_{LTS}$: LTS duration | 6.4 µs ($T_{FFT}$) | 3.2 µs ($T_{FFT}$) |
| $T_{Long}$: Long training sequence duration | 12.8 µs (2 × $T_{LTS}$) | 8 µs ($T_{GI2} + 2 × T_{LTS}$) |
| $\Delta_F$: Freq between sub-carrier | 0.15625 MHz (= 10 MHz/64) | 0.3125 MHz (=20 MHz/64) |
| $T_{FFT}$: T Fast Fourier Transform (IFFT) | 6.4 µs (1/$\Delta F$) | 3.2 µs (1/$\Delta F$) |
| $T_{PREAMBLE}$: PLCP preambule | 32 µs ($T_{Short} + T_{Long}$) | 16 µs ($T_{Short} + T_{Long}$) |
| $T_{SIGNAL}$: OFDM symbol period | 8.0 µs ($T_{GI} + T_{FFT}$) | 4.0 µs ($T_{GI} + T_{FFT}$) |
| $T_{GI1}$: Guard Interval period | 1.6 µs ($T_{FFT}/4$) | 0.8 µs ($T_{FFT}/4$) |
| $T_{GI2}$: Training symbol duration | 3.2 µs ($T_{FFT}/2$) | 1.6 µs ($T_{FFT}/2$) |
| $T_{SYM}$: Symbol interval | 8 µs ($T_{GI} + T_{FFT}$) | 4 µs ($T_{GI} + T_{FFT}$) |
| Time Slot | 13 µs | 9 µs |
| Modulation: BPSK ½, BPSK ¾, QPSK ½, QPSK ¾, 16QAM ½, 16QAM ¾, 64QAM 2/3, 64QAM 3/4 | 3, 4.5, 6, 9, 12, 18, 24, 27 (**MBPS**) | 6, 9, 12, 18, 24, 36, 48, 54 (**MBPS**) |

Table 1: 802.11p/802.11a PHY layer parameters.

## 3. Frame Format:

The PPDU PHY-layer Protocol Data Unit packet is composed of a preamble, signal field and a payload component containing the useful data Fig. 3. The Preamble marks the beginning of the physical frame, select the appropriate antenna and correct the frequency and timing offset. This field is composed of 12 training symbols which are added for temporally synchronization of the reception and for providing a description of the frequency channel behavior. It consists of ten repetitions of a Short Training Symbol STS and two repetitions of a Long Training Symbol LTS [11]. The first 10 STS are short OFDM symbols Table.1.Their roles are the signal detection and the automatic gain control AGC. They also allow the estimation of frequency subcarriers and channel estimation.

A Short Training Symbol consists of 12 subcarriers ± (4, 8, 12, 16, 20, and 24) which are generated directly by the sequence S given by (1):

$S_{-26..26} = \sqrt{\frac{13}{6}} * \{ 0, 0, 1 + j, 0, 0, 0, -1 - j, 0, 0, 0, 1 + j, 0, 0, 0, -1 - j, 0, 0, 0, -1 - j, 0, 0, 0, 1 + j, 0, 0, 0, 0, 0, -1 - j, 0, 0, 0, -1 - j, 0, 0, 0, 1 + j, 0, 0, 0, 1 + j, 0, 0, 0, 1 + j, 0, 0, 0, 1 + j, 0, 0 \}$

(1)

The LTS consists of 53 subcarriers. Their role is essentially to estimate the transmission channel. They are generated directly by applying the IFFT to the following training sequence L (2):

$L_{-26..26} = \{ 1, 1, -1, -1, 1, 1, -1, 1, -1, 1, 1, 1, 1, 1, 1, -1, -1, 1, 1, -1, 1, -1, 1, 1, 1, 1, 0, 1, -1, -1, 1, 1, -1, 1, -1, 1, -1, -1, -1, -1, -1, 1, 1, -1, -1, 1, -1, 1, -1, 1, 1, 1, 1 \}$

(2)

A first Guard Interval GI2 is inserted between the STS and LTS. Its role is to avoid interference between STS and LTS.

The Signal field allows the determining of much crucial transmission information such as the type of modulation to be performed.

It is composed of Rate, Length and Tail fields which determine the coding data rate, the field length and a six bits field set to 0, respectively.

Data field represents the encapsulated data transmitted on the radio channel. It is composed of four fields Service, PSDU, Tail and padding bits fields.

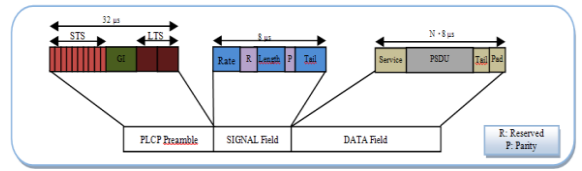

Fig. 3: PPDU PHY Frame format.

## IV. Transmission Architecture:

In order to evaluate the performance of the system, we have used a Matlab SIMULINK tool box to simulate 802.11p PHY layer chain. The encoding process is composed of many complex steps Fig. 4. The following overview describes the details of this procedure convergence.

Since we do not have a relation with the upper layer, the transmitted bits are generated with data source component. These data are scrambled to prevent a long bits sequence that can cause errors during transmission. The scrambler component

generates a sequence of 127 bits based on the function:

$$S(x) = x^7 + x^4 + 1 \qquad (3)$$

To avoid the adverse effect caused by Inter-Symbol Interference ISI and Inter-Carrier Interference ICI, a convolutional encoder, with ½ coding rate, is used on the generated data for error detection and their correction. It can assure adding redundancy to the transmitted bit stream.

In order to obtain upper coding rates, a puncturing element is applied to the output of convolutional encoder resulting in coding rates of R=¾ and R=$2/3$. The puncturing reduces the number of bits to be transmitted and thereby increasing the coding rate. It consists of omitting some of the coded bits on the transmitter side and inserting in their places "zeros" in the convolutional decoder on the receiving side. The puncture model is specified by the binary puncturing vector which corresponds to two bit sequences 1110 for rate R= $2/3$ and 110101 for rate R= ¾. Coded data are interleaved to avoid burst errors due to channel fading. The interleaving process can be divided in two-step permutation in time and in frequency domain. The role of the first permutation is to ensure that two successive bits are never coded in two adjacent subcarriers. The second permutation allows the fact that two successive bits are represented alternately in the most and least significant bits of the used constellation.

The interleaved data are modulated using the phase shift keying (BPSK or QPSK) or the amplitude modulation (16-QAM or 64-QAM). The choice of mapping and coding rate affects the throughput of the transmitted signal which can vary from 3 Mb/s (with BPSK and 1/2 coding rate) to 27 Mb/s (with 64-QAM and 3/4 coding rate). The bit streams are converted into symbols to be transmitted simultaneously. Thus, the OFDM technique converts the serial data stream into several parallel ones. It modulates those data onto orthogonal subcarriers using Inverse Fourier Transform (IFT). The total number of available subcarriers is 64 but only 52 information carriers are used for mapping. Indeed, the 11 guard subcarriers are used on the OFDM spectrum sides to provide separation from adjacent subbands. This step places the complex symbols associated with different constellation points on subcarriers. The IEEE 802.11p uses 52 subcarriers which are exploited as follows: 48 are data subcarriers and 4 are pilot subcarriers. The pilot subcarriers are used to make the detection robust against frequency offsets and phase noise. The pilot symbols are used to estimate the channel and examine the changes made to the transmitted signal. They are inserted in the subcarriers -21, -7, 21, and 7. To transport the data on subcarriers, the OFDM symbols are converted from frequency domain to temporal domain by using IFFT. A second GI1 is inserted before each OFDM symbol in order to avoid the ISI and ICI problems caused by multipath propagation. According to the IEEE 802.11p (Draft 9.0) [11], the GI1 duration is equal to TGI1= $\frac{TFFT}{4}$ Table.1. It consists of copying the end of OFDM symbol in the beginning of the following symbol.

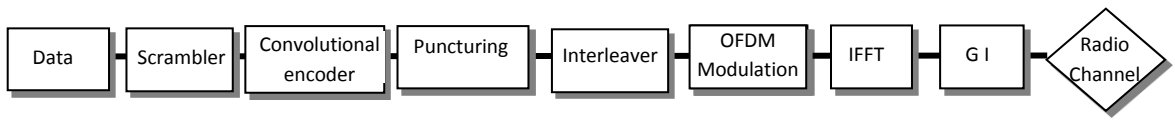

Fig. 4 Transmitter Process for 802.11p Study.

## V. Simulations and Results :

The study of the WAVE standard ability requires focusing on the behavior of its PHY layer depending on a multiple factors such as the mobility of the nodes, the transmission channel type and the transmitted frame size.

*A- Impact of node Mobility on the QoT:*

In this section, we conduct multiple simulations through evaluating the performance of the inter-vehicles communication. In all the cases, we study the variance of the QoT, in particular the SNR effect. Two set of simulations are performed. A comparative study is made for each modulation type depending on whether vehicles are moving in an urban environment or stopped one. The simulation conditions consider the same number of symbols in a transmitted frame (30 symbols). The urban environment is simulated by the Rician radio channel due to the fact that it provides both direct and reflected links between sender and receiver. The speed factor used in the different scenarios is performed according to its dependence on the Doppler Shift variance. The environment selected in the simulations is an urban one as speed with speed varying between 0 Km/h and 50 Km/h. They are all performed for different modulation types and all coding rates.

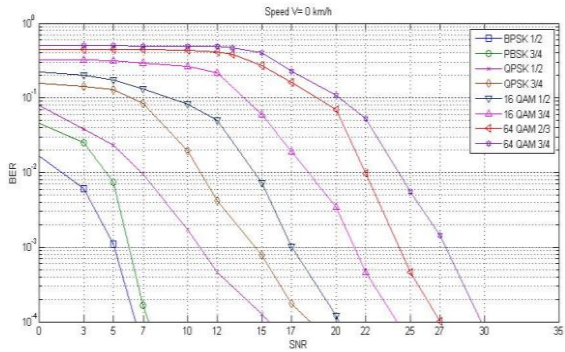
Fig. 5a : BER vs SNR using various modulation, for speed V=0 Km/h.

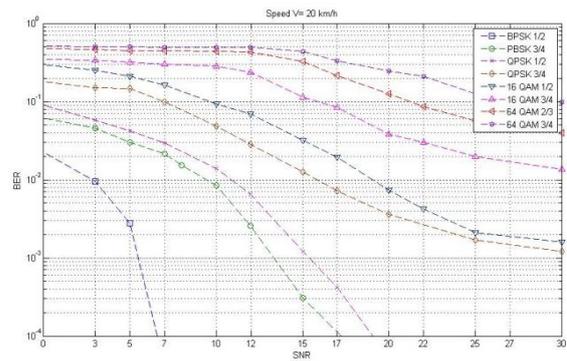
Fig. 5b : BER vs SNR using various modulation, for speed V=20 Km/h.

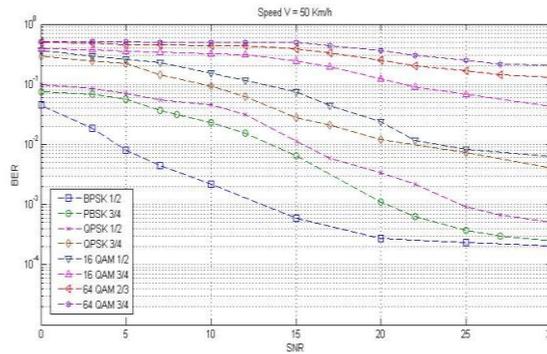
Fig. 5c : BER vs SNR using various modulation, for speed V=50 Km/h.

Fig. 5 (a, b, c) illustrate the dependence between the ratio SNR and the BER: the vertical axes shows the bits error rate while the horizontal axes shows the SNR. We performed our simulation according to different possible modulation. By compiling all curves in plots, we demonstrate that the BER decreases when the transmission quality is better (SNR increase).

Fig 5 shows that the blue curves describe the aspect of the change in BER as a function of SNR for modulation BPSK ½. Blue curves show the best values compared to other modulations. It can be concluded that the BPSK ½ is the best modulation.

By compiling all the figures, we can conclude that the BER is strictly related to the mobility of nodes. In fact, as far as BPSK ½ modulation is concerned, Fig 5a (speed = 0 Km / h), the rate of BER does not exceed the maximum $10^{-3}$. While Fig 5b (speed = 20 Km/h), it reaches $2 \cdot 10^{-2}$.

We found that the more important the speed is, the higher BER will be: Fig 5c (speed = 50 Km/h) shows that the BER is the highest reaching $7 \cdot 10^{-2}$ as compared to the two above values.

We can conclude that the higher the speed is, the more difficult the communication between vehicles becomes due to Doppler Shift effect variation. Consequently, the transmitted signals are attenuated and the BER becomes more important.

It can then be deduced that the optimal quality of the transmission in the 802.11p standard depends on speed variation Fig 6.

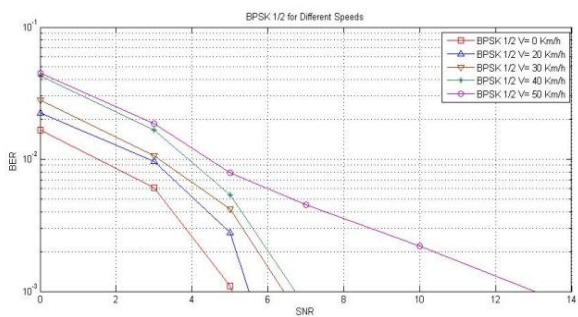
Fig. 6: BER vs SNR using BPSK 1/2, for different speed.

## B- Impact of Channel type on the QoT:

In this section, we perform the BER variation versus the SNR argument focusing on the effect of channel choice. Different channel models are used to evaluate transmission performances.

This study deals with three different transmission scenarios between sender and receiver. We start with considering direct links (AWGN) only, reflected links (Rayleigh Fading) only and both direct and reflected links (Rician fading) between the two parts of transmission.

The vehicle speed is set to 50 km/h and the number of symbols per frame is fixed to 30 symbols. Three scenarios with three types of channels are performed for the three types of coding rate for all modulations type Fig. 7 (a, b, c, d, e, f, g, h).

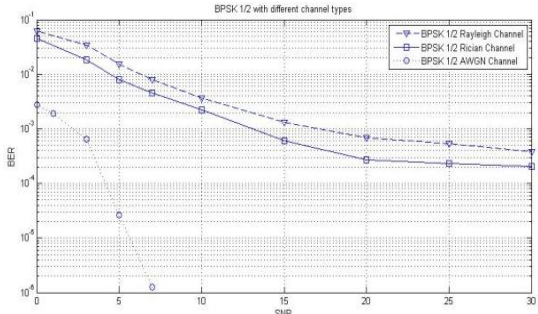

Fig. 7a: BER vs SNR with different channel types for BPSK ½ modulation.

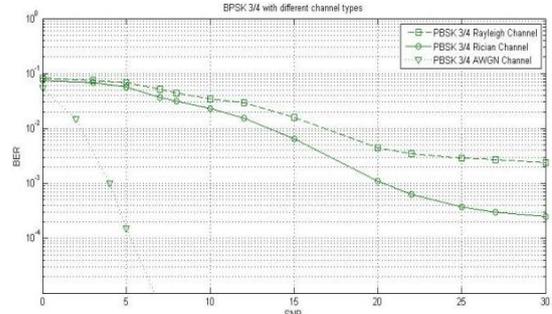

Fig. 7b: BER vs SNR with different channel types for BPSK ¾ modulation

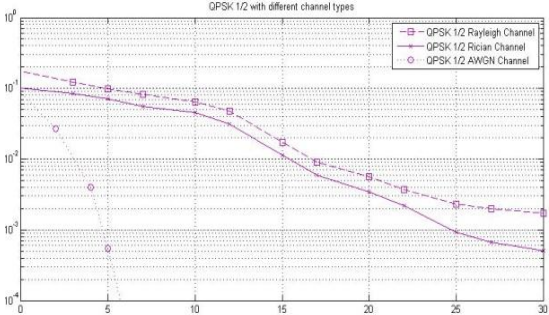

Fig. 7c: BER vs SNR with different channel types for QPSK ½ modulation.

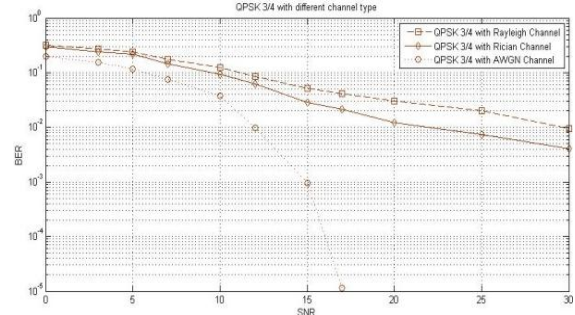

Fig. 7d : BER vs SNR with different channel types for QPSK ¾ modulation

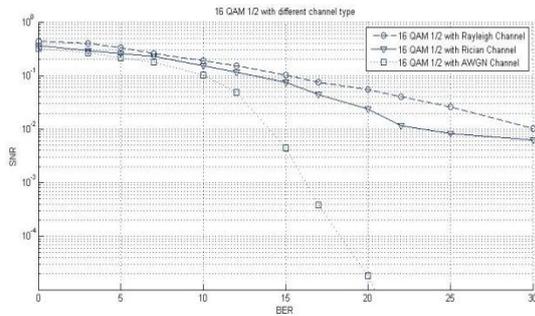

Fig. 7e: BER vs SNR with different channel types for 16 QAM ½ modulation

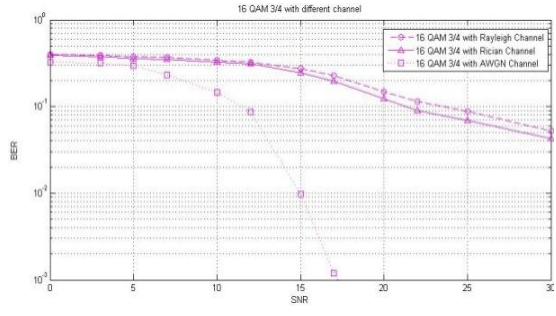

Fig. 7f: BER vs SNR with different channel types for 16 QAM ¾ modulation

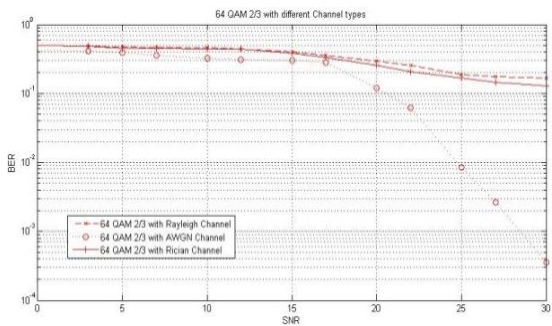

Fig. 7g: BER vs SNR with different channel types for 64 QAM 2/3 modulation

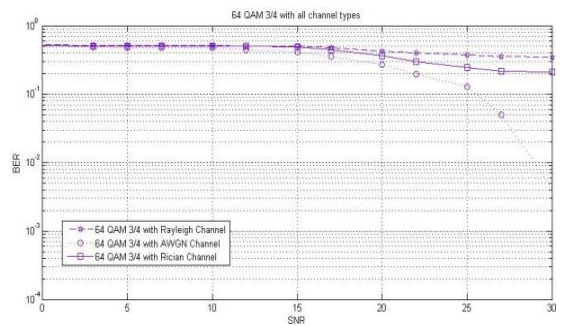

Fig. 7h: BER vs SNR with different channel types for 64 QAM ¾ modulation

The aim of these simulations is to assess the evolution of the BER variance versus the SNR for the three scenarios. The curves evolutions show the relationship between the BER and the quality of the transmission channel. They indicate thus, that BER is higher in the case of Rayleigh channel transmission mode nodes (dashed curves), by comparing to the case of the AWGN channel transmission mode nodes (dotted curve) Fig. 7 (a, b, c, d, e, f, g, h). In practice, we consider a hybrid transmission Rician channel combining the two previous scenarios. The obtained results are logically placed in the middle of the last curves (continuous curve) due to the co-existence of both direct and reflecting links delayed and attenuated.

### C- Impact of transmitted Frame Size on the QoT:

In this section, simulations are performed with the eight modulation types. It considers different number of OFDM symbols in order to study the BER variation Fig. 8. The simulation parameters are set to cope with urban environment conditions. The channel transmission mode is a Rician one while the speed is fixed at 50 km/h. The SNR is fixed at 10 dB.

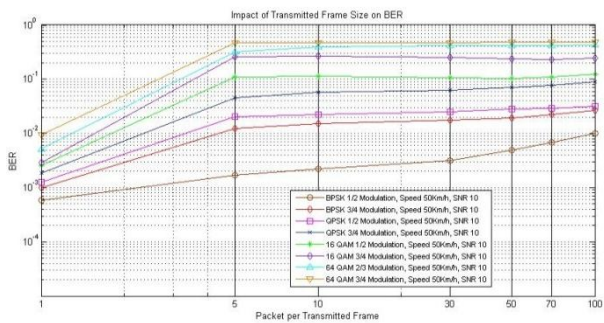

Fig. 8: BER vs Symbol Number per Frame

As the symbol number in the frame increases, the BER grows, too. Consequently, a larger packet length can cause higher error rate. This variation is due to the number of bits to be transmitted each time, hence the following rule.

The more bits to transmit we use, the more important the error risk will be.

The impact of using large number of transmitted bits consists of two main drawbacks: the risks of collision packets and that of congestion of transmission channels.

It is also noted that transmitting with BPSK is better than both QPSK and QAM. It justifies that the Signal field in the OFDM symbol is transmitted with BPSK as it contains a relevant data.

## Conclusion

In this work, we focused on the IEEE 802.11 standard which has been adapted for inter-vehicular communication aiming at providing a low latency and high reliability. A near realistic IEEE.802.11p PHY model for vehicular communication is developed taking account of all the phenomena associated. The BER analysis according to different types of channel reveals that the best channel type is the AWGN but it is still a theoretical case. Hence, the transmission quality closely depends on the SNR variance. The channel type is also one of the most important parameters that may affect the percentage of received erroneous data in transmission. It also reveals that BPSK ½ modulation is the most robust modulation for the different cases. The BER analysis according to Rician channel reveals that both the impact of the node speed and the number of symbols per transmitted frame on the transmission performance. The speed is also one of the most important parameters that affect the transmission reliability. Our future work is twofold. As a first step, a series of simulations in real situations will be performed. In following step, we will attempt to introduce some solutions as to the transmitting as well as receiving parts of the communication.

## References

[1] G. Kiokes A. Amditis N.K. Uzunoglu; "*Simulation-based performance analysis and improvement of orthogonal frequency division multiplexing – 802.11p system for vehicular communications*" 2009, IET Intelligent Transport Systems, pp 429 – 436.

[2] S. Eichler, " *Performance evaluation of the IEEE 802.11p WAVE communication standard*", in Vehicular technology Conference, 2007. VTC-2007 Fall. 2007 IEEE 66th, pp. 2199-2203, 30 oct 2007.

[3] V. Kukshya, H. Krishnan; "*Experimental Measurements and Modeling for Vehicle-to-Vehicle Dedicated Short Range Communication (DSRC) Wireless Channels*" 2007, Vehicular Technology Conference, 2006. VTC-2006 Fall, pp 1 – 5.

[4] A.U. Roberto, A.-M. Gulielmo; "*Wave tutorial, IEEE Commun. Mag*"., 2009, p.47, pp 126-133.

[5] I. Iulia, B.Philippe, B. Xavier, D. Lois, C. Matthieu, D M'hamed; "*Influence of propagation channel modeling on V2X physical layer performance* " Antennas Antennas and Propagation (EuCAP), 2010 Proceedings of the Fourth European Conference, pp 1-5, Barcelona, Spain 2010.

[6] A.M. Vegni and T.D.C. Little, "*A Message Propagation Model for Hybrid Vehicular Communication Protocols*,", in Proc. 2nd Intl. Workshop on Communication Technologies for Vehicles (Nets4Cars), in the 7th IEEE Int. Symp. on Communication Systems, Networks and DSP, (CSNDSP 2010), Newcastle, UK, July 2010

[7] G. Resta, P. Santi, and J. Simon, "*Analysis of multihop emergency message propagation in vehicular ad hoc networks*", in Proc. on the 8th ACM International Symposium on Mobil Ad Hoc Networking and Computing, Montreal, September 09-14, 2007, pp. 140-149.

[8] T. M. Fernandez-Carames, M. Gonzalez-Lopez, and L. Castedo "*FPGA-Based Vehicular Channel Emulator for Real-Time Performance Evaluation of IEEE 802.11p Transceivers*", Volume 2010, Article ID 607467.

[9] L. Bernado, N. Czink, Th. Zemen, P. Belanovie, "*Physical layer simulation results for IEEE 802.11p using vehicular non-stationary channel model*", Communications Workshops (ICC), 2010, p-p 1 – 5.

[10] V. Navda, A. P. Subramanian, K. Dhanasekaran, A. Timm-Giel, and S. Das, "*MobiSteer:using steerable Beam Directional Antenna for Vehicular Network Acess*", in proceedings of the 5th International Conference on Mobile

System, Applications and service (Mobisys '07), New York, NY, USA: ACM. 2007, pp. 192-205.

[11] D.W., Matolak; I., Sen; *"5 GHZ wireless channel characterization for vehicle to vehicle communications",* ilitary Communications Conference, 2005. MILCOM 2005. IEEE, page 3016 - 3022 Vol. 5

[12] Draft amendment to standard for information technology telecommunications and information exchange between systems local and metropolitan area networks specific requirements Part 11: wireless LAN medium access control(MAC) and physical layer specifications Amendment 7: wireless access in vehicular environment, 2007.

[13] IEEE Std 802.11-2007 Part 11: Wireless LAN Medium Access Control (MAC) and Physical Layer (PHY) specifications High-speed Physical Layer in the 5 GHz Band.